\documentclass[aps,pra,showpacs]{revtex4}
\usepackage{amsmath}
\usepackage{graphicx}
\usepackage{dcolumn}
\usepackage{longtable}

\newcommand{\be}{\begin{equation}}
\newcommand{\ee}{\end{equation}}
\newcommand{\bearray}{\begin{eqnarray}}
\newcommand{\eearray}{\end{eqnarray}}
\newcommand{\bse}{\begin{subequations}}
\newcommand{\ese}{\end{subequations}}
 
\newcommand{\half}{\frac{1}{2}}
\newcommand{\third}{\frac{1}{3}}
\newcommand{\fourth}{\frac{1}{4}}
\newcommand{\onebyal}{\frac{1}{\alpha}}
\newcommand{\albypi}{\frac{\alpha}{\pi}}

\begin{document}

\title{Positronium hyperfine splitting at order $m \alpha^7$: light-by-light scattering in the two-photon-exchange channel}

\author{Gregory S. Adkins}
\email[]{gadkins@fandm.edu}
\affiliation{Franklin \& Marshall College, Lancaster, Pennsylvania 17604}
\author{Richard N. Fell}
\affiliation{Brandeis University, Waltham, Massachusetts 01742}

\date{\today}

\begin{abstract}
We have calculated the contribution to the positronium hyperfine splitting at order $m \alpha^7$ of the light-by-light scattering process in the exchange of two photons between the electron and positron.  Our result is $\Delta E =  -0.235355(8) m \alpha^4 \left ( \frac{\alpha}{\pi} \right )^3 = -1.034 kHz$.  As a check of our approach we confirm earlier evaluations of the analogous correction for a bound system (such as muonium) with unequal masses.
\end{abstract}

\pacs{36.10.Dr, 12.20.Ds}

\maketitle


\section{Introduction}
\label{introduction}

Positronium was discovered in 1951 by Martin Deutsch \cite{Deutsch51a}, who immediately measured the three-photon decay rate of the spin-1 variant orthopositronium \cite{Deutsch51b} and, with Everett Dulit, the hyperfine splitting (hfs) \cite{Deutsch51c} with a precision of $15\%$.  Other hfs measurements were performed in the fifties, \cite{Deutsch52,Weinstein54} culminating with 200 ppm results of Deutsch and collaborators \cite{Weinstein55} and of Hughes and collaborators \cite{Hughes57}.  Experimental work resumed on the hfs in the seventies with several measurements by Hughes and collaborators \cite{Theoriot70,Carlson72,Carlson77,Egan77}, leading up to a 3.6 ppm measurement in 1984 \cite{Ritter84}:
\be \Delta E = 203 \, 389.10(74) MHz \quad \text{(3.6 ppm)}. \ee
Also in the seventies, Mills and Bearman \cite{Mills75,Mills83} performed a measurement with similar precision:
\be \Delta E = 203 \, 387.5(1.6) MHz \quad \text{(7.9 ppm)}. \ee
These hfs measurements are all based on observing the Zeeman shift of positronium in a magnetic field and all have systematic errors related to difficulties in determining the magnetic field and on the behavior of positronium in the gas in which it is produced \cite{Isihida12}.  After a lengthy hiatus, new approaches to the measurement of the hfs are being actively developed.  A group based at the University of Tokyo has observed the hyperfine transition directly by stimulating it with high power radiation with several different frequencies in the range 201--206 $GHz$ \cite{Yamazaki12,Namba12}.  The Tokyo group plans to measure a complete transition curve and determine the hfs from the central value.  This group has also improved the quantum oscillation approach \cite{Baryshevski89,Fan96}, using it to perform a new measurement of the hfs \cite{Sasaki11}, and has recently reported a high-precision Zeeman-effect based result \cite{Ishida14}:
\be \Delta E = 203 \, 394.2(1.6)_{\text{stat.}} (1.3)_{\text{sys.}} \quad \text{(10 ppm)} . \ee
 A group based at the University of California, Riverside, has made a preliminary measurement of the positronium hfs using optical transitions between $n=1$ and $n=2$ states and is working to push the method to ppm levels \cite{Cassidy12}.

Theoretical work on the positronium hfs has an even longer history with contributions by many workers.  The theoretical expression for the hfs, showing terms through $O(m \alpha^7)$, can be written as
\be
\Delta E = m \alpha^4 \left \{ C_0 + C_1 \albypi + C_{21} \alpha^2 \ln \left ( \onebyal \right ) + C_{20} \left ( \albypi \right )^2 + C_{32} \frac{\alpha^3}{\pi} \ln^2 \left ( \onebyal \right ) + C_{31} \frac{\alpha^3}{\pi} \ln \left ( \onebyal \right ) +C_{30} \left ( \albypi \right )^3 + \cdots \right \} .
\ee
Terms with and without factors of $\ln \alpha$ are displayed separately so that the $C$ coefficients are pure numbers.  The leading contribution $C_0=7/12$ at $O(m \alpha^4)$ was obtained through the efforts of Pirenne \cite{Pirenne47}, Berestetskii \cite{Berestetskii49}, and Ferrell \cite{Ferrell51} by 1951.  Shortly thereafter, Karplus and Klein obtained the one-loop correction $C_1=-(1/2) \ln 2-8/9$ \cite{Karplus52}.  Early attempts to obtain the two-loop logarithmic corrections, starting in 1970, were incorrect or incomplete because of awkward bound state formalisms and because the origins of the logs were not at first understood.  Many groups contributed to the successful calculation of $C_{21}$\cite{Fulton70,Fulton71,Barbieri73a,Barbieri73b,Owen73,Ng74,Cung76,Barbieri76}.  The correct result, $C_{21}=\frac{5}{24}$, was obtained by Lepage in 1977 and was quickly confirmed \cite{Lepage77,Bodwin78,Caswell79}.  The pure $O(m \alpha^6)$ terms were an even greater challenge, partly because there are so many separate contributions to calculate, partly because some of them required the use of methods that weren't yet developed when work on these terms commenced in 1973.  It took a quarter of a century for work on $C_{20}$ to be successfully completed with contributions from many workers \cite{Barbieri73b,Samuel74,Cung76,Cung77,Cung78a,Cung78b,Caswell78,Cung79,Buchmuller80a,Buchmuller80b,Sapirstein83,Sapirstein84,Caswell86,Adkins88,Adkins93,Karshenboim93a,Zhang94,Hoang97,Adkins97,Pachucki97,Pachucki98,Adkins98,Hoang00,Burichenko01,Adkins02}.  The analytic result $C_{20}=-\frac{53}{32} \zeta(3) + \frac{221}{24} \zeta(2) \ln 2 - \frac{5197}{576} \zeta(2) + \half \ln 2 + \frac{1367}{648}$ was obtained two years after the first correct numerical value \cite{Czarnecki99}.  The leading three-loop log-squared contribution $C_{32} = -\frac{7}{8}$ had already been obtained in 1993 by Karshenboim \cite{Karshenboim93b} and confirmed a few years later \cite{Melnikov99,Pachucki99}.  The single-log term at three-loop order $C_{31} = -\frac{17}{3} \ln 2 + \frac{217}{90}$ was obtained in 2000 \cite{Kniehl00,Melnikov01,Hill01}.  In numerical terms, the theoretical prediction through contribution of $O(m \alpha^7 \ln \alpha)$ is
\be
\Delta E(\text{th}) = 203 \, 391.69 MHz .
\ee
The difference between the theoretical and experimental values for the hfs is $2.59(0.74) MHz$ ($3.5 \sigma$), $4.2(1.6) MHz$ ($2.6 \sigma$), and $-2.5(1.4) MHz$ ($-1.8 \sigma$) for the three measurements, where only the experimental uncertainty is taken into account.  The uncertainty that should be ascribed to the theoretical value is impossible to determine in any precise and universally accepted way.  Kniehl and Penin \cite{Kniehl00} use a value of $0.41 MHz$, based on an analogy with the corresponding $O(m \alpha^7)$ term in the muonium hfs, while Melnikov and Yelkhovsky \cite{Melnikov01} cite $0.16 MHz$, which is half the value of the $O(m \alpha^7 \ln \alpha)$ contribution.  A related estimate of the probable size of the $C_{30}$ contribution comes by comparison with the situation at $O(m \alpha^6)$, where the non-log contribution (from the $C_{20}$ term) with value $-7.33 MHz$ is something like half the size of the  log contribution (from the $C_{21}$ term) with value $19.13 MHz$.  At $O(m \alpha^7)$, the logarithmic terms make contributions of $-0.92 MHz$ ($C_{32}$) and $-0.32 MHz$ ($C_{31}$) for a total of $-1.24 MHz$.  A corresponding estimate of the magnitude of the $C_{30}$ contribution is something like $0.6 MHz$, although $C_{30}$ would have to be $\sim \! 14 \pi^2$ in order to make a contribution of this size.  In fact, sizable contributions can occur at three-loop order.  The one-photon-annihilation contribution at this order was recently found by Baker {\it et al.} to be $C_{30}(1\gamma A) = 49.5(3)$ \cite{Baker14}, making a numerical contribution of $0.217(1) MHz$ to the hfs.  The bulk of this contribution was shown to come from the ``ultrasoft'' scale (energies $\sim m \alpha^2$).  Other ultrasoft contributions were calculated by Marcu \cite{Marcu11} to make a contribution of size $C_{30}{\textrm{(Marcu)}} = 11.0044(10) \pi^2$, which makes a contribution to $\Delta E$ of $0.48 MHz$.  Clearly, having the complete value of $C_{30}$ would be relevant to the comparison between theory and experiment at the present and would be essential for understanding the implications of yet higher precision experiments.  Accordingly, we have initiated a calculation of all contributions to the positronium hfs at $O(m \alpha^7)$. 

We have found the contribution to the hfs of the light-by-light scattering diagrams in the two-photon-exchange channel.  This represents one of many contributions to the full hfs at $O(m \alpha^7)$.  It makes sense to calculate this contribution separately from the rest because it is independent of gauge choice and is independent of the bound state formalism used.  In essence, the light-by-light exchange graphs form a kernel whose expectation value in the positronium bound state gives the corresponding energy shift.  Only the lowest order expressions are needed for the bound state wave functions, which essentially reduce to the wave function at the origin times spin states.
 
In this paper we give a detailed description of our calculation of the positronium hfs due to light-by-light corrections in the two-photon-exchange channel.  We organize the light-by-light diagrams into two classes (ladder and crossed) that we evaluate separately (see Sec.~II).  As a partial check of our procedure, we generalized our calculation to apply to the situation of unequal masses (Sec.~III) so that we could compare with the known muonium hfs results--both in the no-recoil limit and with the inclusion of recoil corrections.  Our results are in accord with known muonium results.


\section{Calculation of the energy shift}
\label{calc}

\begin{figure}
\includegraphics[width=5.0in]{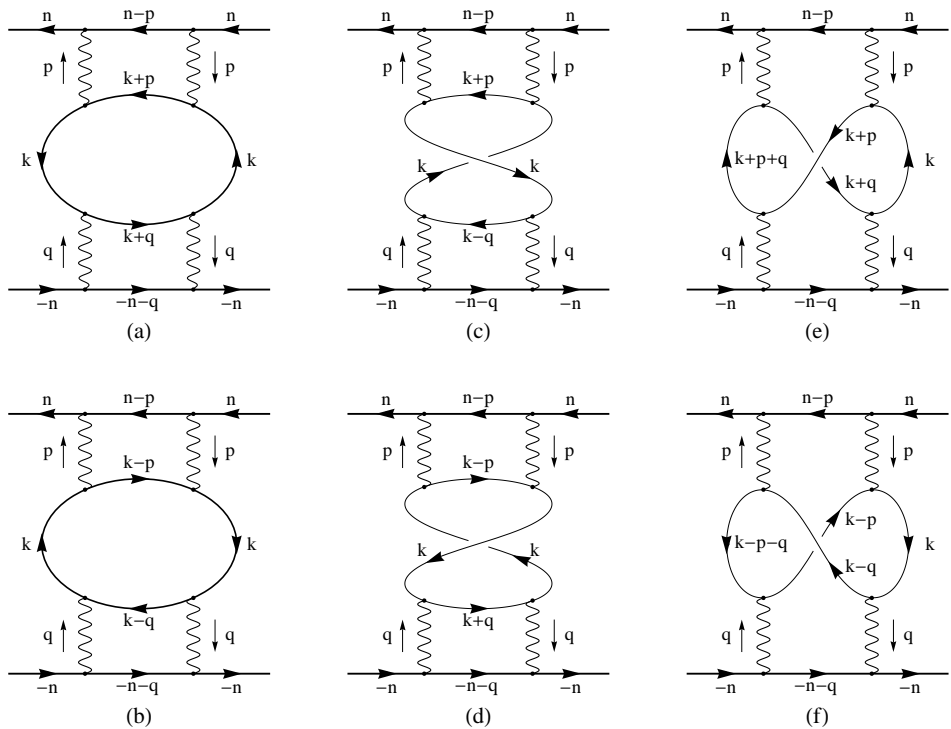}
\caption{\label{fig1} The six light-by-light scattering graphs in the two-photon-exchange channel.  The six graphs represent the six possible ways the four photons can be connected by a closed fermion loop.  The graphs on the bottom row are identical to those on the top except that the light-by-light loop is traversed in the opposite direction.}
\end{figure}

The diagrams that contribute to the light-by-light correction to the two-photon-exchange channel are shown in Fig.~\ref{fig1}.  The electron and positron enter these scattering diagrams from the right and leave towards the left with the electron line on top (using the usual convention that the final state is on the left and the initial state on the right).  The six diagrams represent the six ways that the closed electron loop can connect to four virtual photons.  Momenta are labeled so that these six diagrams are identical outside of the closed electron loop.  The hyperfine splitting from this set of diagrams is finite both in the infrared and ultraviolet.  The order of this three-loop contribution is $m \alpha^7$, with $\alpha^3$ coming from the square of the wave function at the origin and $\alpha^4$ from the interactions at the ends of the four virtual photons.

\begin{figure}
\includegraphics[width=6.4in]{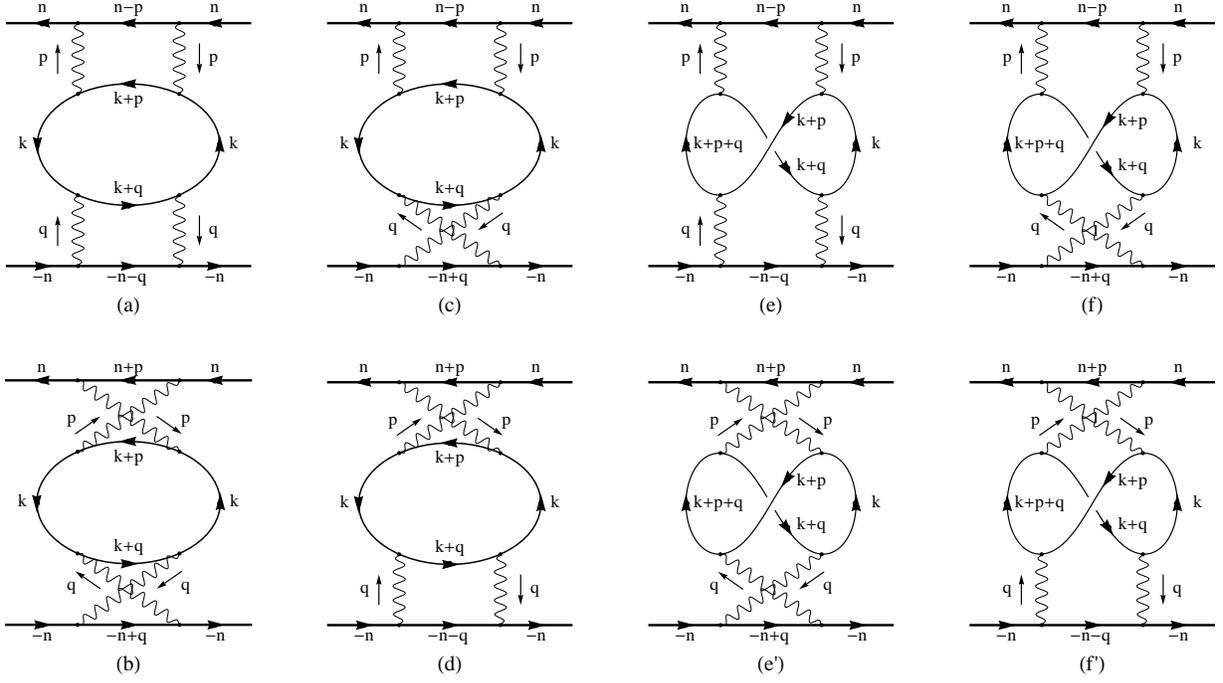}
\caption{\label{fig2} The six light-by-light scattering graphs in the two-photon-exchange channel.  The six graphs represent the six possible ways the four photons can be connected by a closed fermion loop.  The graphs on the bottom row are identical to those on the top except that the light-by-light loop is traversed in the opposite direction.  The leftmost set of four diagrams comprise the 	``ladder'' contribution, while the four diagrams on the right comprise the ``crossed'' contribution.}
\end{figure}

We find it convenient to re-express diagrams of Fig.~\ref{fig1} as the diagrams of Fig.~\ref{fig2}.  Here the diagrams 1(a) 1(b), 1(c), and 1(d) of Fig.~1 are equal to the diagrams 2(a), 2(b), 2(c), 2(d) of Fig.~2, respectively, while the diagram 1(e) is equal to 2(e) and to 2(e$'$), and 1(f) is equal to 2(f) and to 2(f$'$).  The diagrams of Fig.~2 are drawn so that the electron loop has either the form of 2(a)-2(d) or of 2(e)-2(f$'$), completely the same in each set of four diagrams.  It is not difficult to show the equality of the diagrams of Fig.~1 and Fig.~2--one has only to redraw the diagram and redefine some of the momentum variables.  For example, if 1(b) is redrawn by reflecting the electron loop $180^\circ$ around a vertical axis and then redefining $p \rightarrow -p$, $q \rightarrow -q$, one obtains 2(b).  In a similar way, each of the diagrams of Fig.~1 can be shown to equal one or more of the diagrams of Fig.~2.  Following Eides, Karshenboim, and Shelyuto, \cite{Eides91} who calculated a similar contribution to the muonium hfs, we denote the contribution of diagrams 2(a)-2(d) the ``ladder'' contribution and that of 2(e)-2(f$'$) the ``crossed'' contribution.

The light-by-light contribution, as a whole, is ultraviolet finite, infrared safe, and gauge independent.  Each graph by itself has a logarithmic ultraviolet divergence, but when the six permutations are summed the total contribution is ultraviolet finite.  In our calculation we do not perform the fermion loop integrals (having momentum $k$ in Fig.~1 or Fig.~2) first, which would lead to logarithmic divergences graph by graph, instead we integrate over $p$ or $q$ first.  With that order of integration the diagrams are individually ultraviolet finite.  The light-by-light contribution is known to be safe in the infrared having the Euler-Heisenberg effective Lagrangian as its low-energy limit, again when all diagrams are taken as a whole.  We find that the ladder and crossed contributions are individually infrared finite due to cancellation of the leading infrared term when graphs with crossed and uncrossed photons are added together.  Furthermore, there is no threshold sensitivity in the light-by-light contribution--the entering and leaving electron and positron can be taken to be at rest and have energy equal to their rest-energy $m$ (where $m$ is the electron mass).

Because the light-by-light contribution is infrared safe and finite at threshold, we can use practically any bound state formalism to calculate the energy shift that it produces.  In a Bethe-Salpeter-based formalism, the energy shift is just the expectation value of the scattering graphs in a positronium bound state.  The role of the bound-state wave function is essentially to set the relative momentum to zero and to contain the spin state (total spin 0 or spin 1) information.  One such formalism was described in Ref.~\cite{Adkins99}.  The energy shift in this formalism has the form
\be \label{expec_value} \Delta E = i \bar \Psi \delta K \Psi \ee
where $\delta K$ is the light-by-light interaction kernel illustrated in Figs.~1 and 2 and $\Psi$ is the wave function.  This is only the leading order contribution from light-by-light scattering as higher order contributions containing a light-by-light sub-part also exist.  At leading order the wave function can be approximated as
\be \Psi \rightarrow (2 \pi)^4 \delta^4 (\ell) \phi_0 \begin{pmatrix} 0 & \chi \\ 0 & 0 \end{pmatrix}, \quad \bar \Psi^T \rightarrow (2 \pi)^4 \delta^4 (\ell) \phi_0 \begin{pmatrix} 0 & 0 \\ \chi^\dagger & 0 \end{pmatrix} \ee
where $\ell$ is the relative momentum, $\phi_0=\sqrt{(m/2)^3 \alpha^3/\pi}$ is the wave function at spatial contact, and $\chi$ is the Pauli $2 \times 2$ spin state: $\chi_{0 0}=1/\sqrt{2}$ for the parapositronium ($S=0$) state and $\chi_{1,m}=\vec \sigma \cdot \hat \epsilon_m/\sqrt{2}$ for the orthopositronium ($S=1$) state, where $\hat \epsilon_m$ is the normalized orthopositronium spin vector.  The explicit interpretation of the expectation value \eqref{expec_value} for a situation involving a particle-antiparticle bound state is
\be \label{expec_trace} \Delta E = i \mathrm{tr} \left [ \bar \Psi^T A \Psi B \right ] \ee
where $A$ is the part of the interaction kernel involving electron line factors and $B$ is the corresponding positron line factors.  (Other factors that are not specifically associated with either line, such as the light-by-light scattering tensor itself, can be associated with either $A$ or $B$.)  So, for example, the first of the ladder diagrams contributes
\bearray
\Delta E_a &=& i \phi_0^2 \int \frac{d^4 p}{(2 \pi)^4} \frac{d^4 k}{(2 \pi)^4} \frac{d^4 q}{(2 \pi)^4} \mathrm{tr} \Bigl [ \begin{pmatrix} 0 & 0 \\ \chi^\dagger & 0 \end{pmatrix} (-i e \gamma^\alpha) \frac{i}{\gamma(P/2-p)-m} (-i e \gamma^\beta) \cr
&\times& \begin{pmatrix} 0 & \chi \\ 0 & 0 \end{pmatrix} (-i e \gamma^\mu) \frac{i}{\gamma(-P/2-q)-m} (-i e \gamma^\nu) \Bigr ]
\left ( \frac{-i}{p^2} \right )^2 \left ( \frac{-i}{q^2} \right )^2 \cr
&\times&  (-1) \; \mathrm{tr} \Bigl [ (-i e \gamma_\mu) \frac{i}{\gamma(k+q)-m} (-i e \gamma_\nu) \frac{i}{\gamma k-m} (-i e \gamma_\alpha) \frac{i}{\gamma(k+p)-m} (-i e \gamma_\beta) \frac{i}{\gamma k-m} \Bigr ] ,
\eearray
where $P=(2m,\vec 0 \, )$ is the rest frame positronium energy-momentum vector and the factor of $-1$ in front of the light-by-light trace is the  sign that arises whenever there is a closed fermion loop.  We use $e^2=4 \pi \alpha$ to rewrite this contribution as
\bearray \label{deltaEa_2}
\Delta E_a &=& \frac{m \alpha^7}{128 \pi^3} \int \frac{d^4 p}{i \pi^2} \frac{d^4 k}{i \pi^2} \frac{d^4 q}{i \pi^2}  \mathrm{tr} \Bigl [ \begin{pmatrix} 0 & 0 \\ \chi^\dagger & 0 \end{pmatrix} \gamma^\alpha \bigl ( \gamma(n-p)+1 \bigr ) \gamma^\beta \begin{pmatrix} 0 & \chi \\ 0 & 0 \end{pmatrix} \gamma^\mu \bigl ( \gamma(-n-q)+1 \bigr ) \gamma^\nu \Bigr ] \cr 
&\times& \mathrm{tr} \Bigl [ \gamma_\mu \bigl ( \gamma (k+q)+1 \bigr ) \gamma_\nu \bigl ( \gamma k + 1 \bigr ) \gamma_\alpha \bigl ( \gamma (k+p)+1 \bigr ) \gamma_\beta \bigl ( \gamma k + 1 \bigr ) \Bigr ] \cr
&\times& \Bigl [ (-p^2)^2 (-q^2)^2 (-(n-p)^2+1) (-(-n-q)^2+1) (-k^2+1)^2 (-(k+q)^2+1) (-(k+p)^2+1) \Bigr ]^{-1} ,
\eearray
where we have scaled all momenta by the electron mass $m$ and have defined $n=P/(2m)=(1,\vec 0 \,)$.  Implicit in each propagator denominator is a term $-i \epsilon$ with $\epsilon \rightarrow 0^+$ so that $-k^2+1$ really means $-k^2+1-i \epsilon$, for example.

Charge conjugation arguments may be used to show that the top three diagrams of Fig.~1: 1(a), 1(c), and 1(e), are equal to the bottom three diagrams: 1(b), 1(d), and 1(f), respectively.  The procedure is to take the transpose of the fermion loop trace (which doesn't change its value), use the charge conjugation matrix $C=i \gamma^2 \gamma^0$ to rewrite transposed gamma matrices according to $\gamma^{\mu T} = -C^{-1} \gamma^\mu C$, remove the $C$ matrices using $C^{-1}C=1$, and redefine the momentum $k \rightarrow -k$.  Only two of the ladder diagrams are independent, and there is only one independent crossed diagram.  However, we find it convenient to keep all four ladder diagrams and all four crossed diagrams as shown in Fig.~2 because having all four of each makes the favorable infrared behavior manifest.

Now we will take a closer look at the electron-positron line factor as it contributes to the o-Ps--p-Ps energy difference.  In particular, we will look at the trace
\be \label{line_trace_1}
T \equiv \mathrm{tr} \Bigl [ \begin{pmatrix} 0 & 0 \\ \chi^\dagger & 0 \end{pmatrix} \gamma^\alpha \bigl ( \gamma(n-p)+1 \bigr ) \gamma^\beta \begin{pmatrix} 0 & \chi \\ 0 & 0 \end{pmatrix} \gamma^\mu \bigl ( \gamma(-n-q)+1 \bigr ) \gamma^\nu \Bigr ] .
\ee
We notice that the positronium spin matrices can be written in terms of positive and negative energy projection operators $\Lambda_\pm = \half (1 \pm \gamma^0 )$ and the appropriate spin factors as
\be
\begin{pmatrix} 0 & \chi \cr 0 & 0 \end{pmatrix} = \frac{1}{\sqrt{2}} \Lambda_+ \Bigl \{ \vec \gamma \cdot \hat \epsilon_m , \gamma_5\Bigr \} \Lambda_- , \quad
\begin{pmatrix} 0 & 0 \cr \chi^\dagger & 0 \end{pmatrix} = \frac{1}{\sqrt{2}} \Lambda_- \Bigl \{ - \vec \gamma \cdot \hat \epsilon^*_m , \gamma_5\Bigr \} \Lambda_+ 
\ee
for $\{$o-Ps, p-Ps$\}$, respectively.  The corresponding traces simplify to
\be
T_{\{\text{o-Ps, p-Ps} \}} = \half \mathrm{tr} \Bigl [ \Lambda_- \Bigl \{ - \vec \gamma \cdot \hat \epsilon^*_m , \gamma_5\Bigr \} \Lambda_+  \Bigl ( \gamma^\alpha\gamma p \gamma^\beta - 2 n^\alpha n^\beta \Bigr ) \Lambda_+ \Bigl \{ \vec \gamma \cdot \hat \epsilon_m , \gamma_5\Bigr \} \Lambda_- \Bigl ( \gamma^\mu \gamma q \gamma^\nu - 2 n^\mu n^\nu\Bigr ) \Bigr ] .
\ee
Spatial symmetry allows us to compute the o-Ps energy using any direction for the spin $\hat \epsilon_m$.  In fact, we average over the three coordinate directions since the three contributions are equal.  The trace becomes
\be
T_{\{\text{o-Ps, p-Ps} \}} = \half \mathrm{tr} \Bigl [ \Lambda_- \Bigl \{ \gamma^\lambda , \gamma_5\Bigr \} \Lambda_+  \Bigl ( \gamma^\alpha\gamma p \gamma^\beta - 2 n^\alpha n^\beta \Bigr ) \Lambda_+ \Bigl \{ \third \gamma_\lambda , \gamma_5\Bigr \} \Lambda_- \Bigl ( \gamma^\mu \gamma q \gamma^\nu - 2 n^\mu n^\nu\Bigr ) \Bigr ] .
\ee
since the $\lambda=0$ contribution vanishes.  In the hfs energy difference o-Ps minus p-Ps, the $n^\alpha n^\beta$ and $n^\mu n^\nu$ terms cancel, leaving
\be
\Delta T = \frac{1}{6} \mathrm{tr} \Bigl [ \Bigl ( \Lambda_- \gamma^\lambda \Lambda_+ \gamma^\alpha\gamma p \gamma^\beta \Lambda_+ \gamma_\lambda \Lambda_- \gamma^\mu \gamma q \gamma^\nu \Bigr )
-3 \Bigl ( \Lambda_- \gamma_5 \Lambda_+ \gamma^\alpha\gamma p \gamma^\beta \Lambda_+ \gamma_5 \Lambda_- \gamma^\mu \gamma q \gamma^\nu \Bigr ) \Bigr ] .
\ee

We note that the hyperfine energy difference is better behaved in the infrared than the energy for either state separately.  Further improvement in the infrared is made manifest when all combinations of uncrossed and crossed photons are included together.  The only difference between the diagrams 2(a) and 2(d), for example, is in the order of the photons entering the electron line and in the sign of $p$ in the electron line (numerator and denominator both).  The difference inside the trace is that while 2(a) has a term $\Lambda_+ \gamma^\alpha \gamma p \gamma^\beta \Lambda_+$, 2(d) has instead $\Lambda_+ \gamma^\beta (-\gamma p) \gamma^\alpha \Lambda_+ = \Lambda_+ \left ( \gamma^\alpha \gamma p \gamma^\beta - 2n^\alpha p^\beta-2 n^\beta p^\alpha + 2 g^{\alpha \beta} p^0 \right ) \Lambda_+ \rightarrow \Lambda_+ \gamma^\alpha \gamma p \gamma^\beta \Lambda_+$ since the omitted terms vanish in the hfs energy difference.  As a consequence, the only effect of adding 2(d) to 2(a) is to replace the electron line denominator $(-p^2+2 p \cdot n)^{-1}$ by
\be
\frac{1}{-p^2+2 p \cdot n} + \frac{1}{-p^2-2 p \cdot n} = \frac{-2p^2}{(-p^2+2 p \cdot n)(-p^2-2 p \cdot n)}
= \int_0^1 du \frac{-2 p^2}{(-p^2+2 p \cdot n \bar u )^2} ,
\ee
with $\bar u = 1-2u$.  An analogous term arises when we combine the uncrossed and crossed photons entering the positron line.  Where \eqref{deltaEa_2} had the apparently problematic infrared behavior
\be \int d^4 p \frac{\gamma (n-p) + 1}{(-p^2)^2 (-p^2+2 p \cdot n)}  \ee
for, say, the $p$ integral,
the uncrossed plus crossed hfs combination has the more agreeable form
\be -2 \int_0^1 du \int d^4 p \frac{\gamma p}{(-p^2) (-p^2+2 p \cdot n \bar u)^2}  . \label{combined_form} \ee

By combining terms as discussed above, the complete ladder contribution (from diagrams 2(a)-2(d)) can be written as
\bearray \label{ladder_int}
\Delta E_L &=& \frac{m \alpha^7}{12 \pi^3} \int_0^1 du \int_0^1 dv \int \frac{d^4 p}{i \pi^2} \frac{d^4 k}{i \pi^2} \frac{d^4 q}{i \pi^2} N_L \Bigl [ (-p^2) (-p^2+2 p \cdot n \bar u)^2 (-q^2)  (-q^2+2 q \cdot n \bar v)^2 \cr
&\hbox{}& \quad \quad \times (-k^2+1)^2 (-(k+q)^2+1) (-(k+p)^2+1) \Bigr ]^{-1} ,
\eearray
where $\bar u=1-2u$, $\bar v=1-2v$, and $N_L$ is the trace factor
\bearray
N_L &=& \fourth \mathrm{tr} \Bigl [ \Bigl ( \Lambda_- \gamma^\lambda \Lambda_+ \gamma^\alpha\gamma p \gamma^\beta \Lambda_+ \gamma_\lambda \Lambda_- \gamma^\mu \gamma q \gamma^\nu \Bigr )
-3 \Bigl ( \Lambda_- \gamma_5 \Lambda_+ \gamma^\alpha\gamma p \gamma^\beta \Lambda_+ \gamma_5 \Lambda_- \gamma^\mu \gamma q \gamma^\nu \Bigr ) \Bigr ] \cr
&\times& \; \; \fourth \mathrm{tr} \Bigl [ \gamma_\mu \bigl ( \gamma (k+q)+1 \bigr ) \gamma_\nu \bigl ( \gamma k + 1 \bigr ) \gamma_\alpha \bigl ( \gamma (k+p)+1 \bigr ) \gamma_\beta \bigl ( \gamma k + 1 \bigr ) \Bigr ] .
\eearray
We perform the momentum integrals through the introduction of six additional parameters, performing the integrals in the order $p$, $q$, $k$.  Specifically, we associate $1-z$ with $-p^2$, $z(1-x)$ with $-p^2+2 p \cdot n \bar u$, and $z x$ with $-(p+k)^2+1$ for the $p$ integral, $1-y$ with $(-q)^2$, $y(1-w)$ with $-q^2+2 q \cdot n \bar v$, and $y w$ with $-(q+k)^2+1$ for the $q$ integral, and finally $1-t$ with $-k^2+1$, $t(1-s)$ with $-k^2 + 2k \cdot n B_p + C_p$, and $t s$ with $-k^2 + 2 k \cdot n B_q + C_q$ for the $k$ integral, where $a_p=z x (1-z x)$, $B_p = -z^2 x (1-x) \bar u/a_p$, $C_p=z (x+z(1-x)^2 \bar u^2)/a_p$, $a_q=y w (1-y w)$, $B_q = -y^2 w (1-w) \bar v/a_q$, and $C_q=y (w+y(1-w)^2 \bar v^2)/a_q$.  The ladder contribution then has the form
\bearray \label{ladder_integral}
\Delta E_L &=& \frac{m \alpha^7}{12 \pi^3} \int du \, dv \, dz \, dx \, dy \, dw \, dt \, ds \, z^2 (1-x) y^2 (1-w) \frac{t(1-t)}{a_p a_q} \cr
&\hbox{}& \quad \quad \times \Bigl \{ N_0 \frac{6}{\Delta^4} - \half N_1 \frac{2}{\Delta^3} + \fourth N_2 \frac{1}{\Delta^2} - \frac{1}{8} N_3 \frac{1}{\Delta} \Bigr \} ,
\eearray
where
\be \Delta = t^2 \bigl ( (1-s) B_p + s B_q \bigr )^2 + t \bigl ( (1-s) C_p + s C_q \bigr ) + (1-t) , \ee
and
\bse
\bearray
N_0 &=& \frac{t^2 s (1-s)}{a_p a_q} N_{000} , \\
N_1 &=& \frac{t^2 s (1-s)}{a_p a_q} N_{001} + \frac{t s}{a_q} N_{100}+ \frac{t(1-s)}{a_p} N_{010} , \\
N_2 &=& \frac{t^2 s (1-s)}{a_p a_q} N_{002} + \frac{t s}{a_q} N_{101}+ \frac{t(1-s)}{a_p} N_{011} + N_{110} , \\
N_3 &=& \frac{t^2 s (1-s)}{a_p a_q} N_{003} + \frac{t s}{a_q} N_{102}+ \frac{t(1-s)}{a_p} N_{012} + N_{111} .
\eearray
\ese
Here $N_{\ell m n}$ represents the trace term $N_L$ with first $\ell$ contractions applied to $p$ with any remaining $p^\mu$ factors replaced by $Q_p^\mu=-z x k^\mu + z(1-x) \bar u n^\mu$, then $m$ contractions applied to $q$ with any remaining $q^\mu$ factors replaced by $Q_q^\mu=-y w k^\mu + y (1-w) \bar v n^\mu$, then $n$ contractions applied to $k$ with any remaining $k^\mu$ factors replaced by $Q_k^\mu = t \bigl ( (1-s) B_p+s B_q \bigr ) n^\mu$.  (The ``contraction'' process, over $p$ as an example, consists of finding all pairs of momentum vectors, say $p^\mu p^\nu$, replacing the pair by the metric $\eta^{\mu \nu}$, summing over all such pairs, and replacing any remaining $p^\mu$ factors by $Q_p^\mu$.)  The traces and contractions were performed using Reduce \cite{Hearn04}.  There is one further complication before turning the expression over for numerical integration.  It turns out that the infrared sensitivity lurking near $\bar u = \bar v = 0$ is canceled {\em pointwise}.  That is, there is a cancellation between the region just below $\bar u=0$ and the region just above $\bar u=0$, and likewise near $\bar v=0$.  The numerical integration routine that we use, Vegas \cite{Lepage78}, chooses points on both sides of $\bar u = \bar v = 0$, but not necessarily symmetrically.  We enforce the pointwise cancellation around $\bar u = \bar v = 0$ by replacing our integrand $f(\bar u,\bar v)$ (where only the dependence on $\bar u$ and $\bar v$ is displayed), by $\left ( f(\bar u,\bar v) + f(-\bar u,\bar v) + f(\bar u,- \bar v) + f(- \bar u,- \bar v) \right )/4$.  The numerical result for the ladder contribution, obtained using Vegas with 100 iterations of $2\times 10^9$ points each, was
\be \Delta E_L = \frac{m \alpha^7}{\pi^3} \bigl ( 1.483609(5) \bigr ) . \label{ladder_result} \ee

The crossed contribution (from diagrams 2(e)-2(f$'$)), has a form that is analogous to \eqref{ladder_int}.  It is
\bearray \label{crossed_int}
\Delta E_C &=& \frac{m \alpha^7}{24 \pi^3} \int_0^1 du \int_0^1 dv \int \frac{d^4 p}{i \pi^2} \frac{d^4 k}{i \pi^2} \frac{d^4 q}{i \pi^2} N_C \Bigl [ (-p^2) (-p^2+2 p \cdot n \bar u)^2 (-q^2)  (-q^2+2 q \cdot n \bar v)^2 \cr
&\hbox{}& \quad \quad \times (-k^2+1) (-(k+q)^2+1) (-(k+p)^2+1) (-(k+p+q)^2+1) \Bigr ]^{-1} ,
\eearray
where the trace factor $N_C$ is
\bearray
N_C &=& \fourth \mathrm{tr} \Bigl [ \Bigl ( \Lambda_- \gamma^\lambda \Lambda_+ \gamma^\alpha\gamma p \gamma^\beta \Lambda_+ \gamma_\lambda \Lambda_- \gamma^\mu \gamma q \gamma^\nu \Bigr )
-3 \Bigl ( \Lambda_- \gamma_5 \Lambda_+ \gamma^\alpha\gamma p \gamma^\beta \Lambda_+ \gamma_5 \Lambda_- \gamma^\mu \gamma q \gamma^\nu \Bigr ) \Bigr ] \cr
&\times& \; \; \fourth \mathrm{tr} \Bigl [ \gamma_\mu \bigl ( \gamma (k+q)+1 \bigr ) \gamma_\alpha \bigl ( \gamma (k+p+q) + 1 \bigr ) \gamma_\nu \bigl ( \gamma (k+p)+1 \bigr ) \gamma_\beta \bigl ( \gamma k + 1 \bigr ) \Bigr ] .
\eearray
The only differences between the ladder contribution of \eqref{ladder_int} and the crossed contribution here are the replacement of one electron propagator $-i/(\gamma k-m)$ by $-i/(\gamma (k+p+q)-m)$ in the fermion loop, a different ordering of indices in the fermion loop, and the extra factor of $1/2$ due to the fact that the four crossed graphs come from two graphs of Fig.~1 instead of four.  There will be a total of nine parameters instead of eight because of the extra propagator factor.  In fact, we find that
\bearray \label{crossed_integral}
\Delta E_C &=& \frac{m \alpha^7}{24 \pi^3} \int du \, dv \, dz \, dx \, dy \, dw \, dr \, ds \, dt \, z^3 x (1-x) w^4 r^2 (1-r) s\frac{t^2}{a_p^2 a_q^3} \cr
&\hbox{}& \quad \quad \times \Bigl \{ N_0 \frac{6}{\Delta^4} - \half N_1 \frac{2}{\Delta^3} + \fourth N_2 \frac{1}{\Delta^2} - \frac{1}{8} N_3 \frac{1}{\Delta} \Bigr \} .
\eearray
We have performed the $p$ integral first, associating $1-z$ with $-p^2$, $z(1-x)$ with $-p^2+2 p \cdot n \bar u$, $z x (1-y)$ with $-(p+k)^2+1$, and $z x y$ with $-(p+k+q)^2+1$.  The $p$ contractions were performed with the replacement $p^\mu \rightarrow Q_p^\mu= v_{pq} q^\mu + v_{pk} k^\mu + v_{pn} n^\mu$ with $v_{pq}=-z x y$, $v_{pk}=-z x$, $v_{pn}=z(1-x)\bar u$ after all replacements $p^\mu p^\nu \rightarrow \eta^{\mu \nu}$ were done.  We define $a_p=z x y (1-z x y)$, $B_p^\mu = ((1+v_{pk}) k^\mu + v_{pn} n^\mu)(v_{pq}/a_p)$, and $C_p=\left ( (1+v_{pk}) k^2 + v_{pn} 2 k \cdot n + v_{pn}^2/v_{pk} - 1 \right )(v_{pk}/a_p)$.  We next did the $q$ integral, correlating $1-w$ with $-q^2$, $w(1-r)$ with $-q^2+2 q \cdot n \bar v$, $w r (1-s)$ with $-(q+k)^2+1$, and $w r s$ with $\Delta_p=Q_p^2 + z x (1-y) (-k^2+1) + z x y (-(k+q)^2+1)=a_p \left ( -q^2+2 q \cdot B_p + C_p \right )$.  The $q$ contractions were performed with the replacement $q^\mu \rightarrow Q_q^\mu = v_{qk} k^\mu+ v_{qn} n^\mu$ and the definitions $v_{qk}=-w r (1-s) + w r s v_{pq} (1+v_{pk})/a_p$, $v_{qn}=w (1-r) \bar v + w r s v_{pq} v_{pn}/a_p$, $a_q=w r (1-s) - v_{qk}^2 - w r s v_{pk} (1+v_{pk})/a_p$, $B_q^\mu = \left ( v_{qk} v_{qn} + w r s v_{pk} v_{pn} /a_p \right ) n^\mu/a_q$, and $C_q = \left ( w r (1-s) + v_{qn}^2 + w r s (v_{pn}^2-v_{pk})/a_p \right )/a_q$.  Finally, the $k$ integral was done using the associations $1-t$ with $-k^2+1$ and $t$ with $\Delta_q=Q_q^2+w r (1-s)(-k^2+1)+w r s C_p = a_q \left ( -k^2+2 k \cdot B_q + C_q \right )$.  The replacement used in the $k$ contractions was $k^\mu \rightarrow Q_k^\mu = t B_q^\mu$, and the $\Delta$ factor entering \eqref{crossed_integral} is
\be \Delta = t^2 B_q^2 + t C_q + (1-t) . \ee
The $N_i$ terms for \eqref{crossed_integral} are
\bse
\bearray
N_0 &=& \frac{w r s t^2}{a_p a_q^2} N_{000} , \\
N_1 &=& \frac{w r s t^2}{a_p a_q^2} N_{001} + \frac{w r s t}{a_p a_q} N_{010}+ \frac{t}{a_q} N_{100} , \\
N_2 &=& \frac{w r s t^2}{a_p a_q^2} N_{002} + \frac{w r s t}{a_p a_q} N_{011}+ \frac{t}{a_q} N_{101} + \frac{w r s}{a_p} N_{020} + N_{110} , \\
N_3 &=& \frac{w r s t^2}{a_p a_q^2} N_{003} + \frac{w r s t}{a_p a_q} N_{012}+ \frac{t}{a_q} N_{102} + \frac{w r s}{a_p} N_{021} + N_{111} .
\eearray
\ese
Here $N_{\ell m n}$ is obtained from $N_C$ by applying first $\ell$ contractions to $p$, then $m$ contractions to $q$, and finally $n$ contractions to $k$.  The Vegas result for the crossed contribution at $100 \times 2G$ functional evaluations was
\be \Delta E_C = \frac{m \alpha^7}{\pi^3} \bigl ( -1.718964(6) \bigr ) . \label{crossed_result} \ee


\section{Discussion and Comparison with Muonium hfs Results}
\label{results}

The total contribution of light-by-light in the two-photon-exchange channel to the positronium hfs is the sum  of \eqref{ladder_result} and \eqref{crossed_result}:
\be \Delta E_C = \bigl ( -0.235355(8) \bigr ) \frac{m \alpha^7}{\pi^3} = -1.034 kHz . \label{total_result} \ee
This result is numerically small relative to the expected precision of present measurements, partially due to the significant cancellation between the ladder and crossed contributions.  As shown by Marcu \cite{Marcu11} and by Baker {\it et al.} \cite{Baker14}, other contributions at this same three-loop ($m \alpha^7$) order can be much larger.  In any case, the light-by-light contribution computed here is one of many that will need to be evaluated in order to obtain the full three-loop correction.

We took several measures in order to ensure the correctness of our result.  First, we both did the calculations independently and obtained consistent values.  We also checked our normalization by re-computing the hfs corrections for muonium and, in the non-recoil limit, hydrogen.  We briefly describe these calculations here.

For our purposes here, the only distinction between muonium and positronium is the mass difference between the positron and muon.  This mass difference enters the calculation in two places: in the wave function at contact, which now involves the reduced mass: $\phi_0^2 \rightarrow (m_r \alpha)^3/\pi$, where $m_r = (1/m_e+1/m_\mu)^{-1}$; and in the muon line factor.  The spin part of the wave function is unchanged and we are still dealing with an electron light-by-light loop.  The muon propagator $i \left (\gamma (-m_\mu n - q ) - m_\mu \right )^{-1}$, in combination with $i \left (\gamma (-m_\mu n + q ) - m_\mu \right )^{-1}$ from the diagram with crossed photons, leads to
\be -2 \int_0^1 dv \int d^4 q \frac{\gamma q}{(-q^2) (-q^2+2 q \cdot m_\mu n \bar v)^2}  \ee
by the same analysis that produced \eqref{combined_form}.  After factoring out the electron mass the muon denominator factor takes the form $(-q^2+2 q \cdot n \kappa \bar v)^2$ where $\kappa=m_\mu/m_e$ is the muon--electron mass ratio.  In order to compare our muonium hfs results with prior work, we pull out $E_F \left (\alpha^3/\pi \right )$ instead of our $m_e \alpha^7/\pi^3$, where
\be E_F \equiv \frac{m \alpha^4}{3} \left ( \frac{m_e}{m_\mu} \right ) \left ( \frac{2 m_r}{m_e} \right )^3 \ee
is the Fermi splitting.  The muonium hfs contribution becomes $\Delta E = E_F \left (\alpha^3/\pi \right ) \bar I$ where the integral $\bar I$ here differs from the $I$ of \eqref{ladder_integral} or \eqref{crossed_integral} in only two ways: $\bar I = (3 \kappa/\pi^2) I(\bar v \rightarrow \kappa \bar v)$.  That is, $\bar I$ differs from $I$ by the replacement of $\bar v$ by $\kappa \bar v$ and by an overall factor of $3 \kappa/\pi^2$.  A direct evaluation at the physical value $\kappa=m_\mu/m_e=209.7682843(52)$ \cite{Mohr12} for muonium yields
\bearray \Delta E &=& \left \{ 0.809101(20) - 1.244191(23) \right \} E_F \frac{\alpha^3}{\pi} \cr &=& -0.435090(31) E_F \frac{\alpha^3}{\pi}, \eearray
with ladder and crossed contributions displayed separately.  However, earlier workers did not evaluate $\bar I$ for muonium directly, but instead performed an expansion about the no-recoil (infinite-mass muon) limit.  Their result, as an expansion in $1/\kappa$, has the form \cite{Eides91,Karshenboim92,Kinoshita94,Kinoshita96,Eides13}
\be \Delta E = E_F \frac{\alpha^3}{\pi} \left \{ -0.472514(1) + \frac{1}{\pi^2 \kappa} \left [ \frac{9}{4} \ln^2 \kappa + \left ( -3 \zeta(3) -4 \zeta(2) + \frac{91}{8} \right ) \ln \kappa + C_{10} \right ] + \cdots \right \} . \label{expansion}
\ee
We evaluated $\bar I$ at various values of $\kappa$ as displayed in Table~\ref{table1}.  Our fits for the no-recoil limit and the logarithmic terms are in complete accord with the values shown in \eqref{expansion}.  In addition, we obtain a fit to $C_{10}$: $C_{10} \approx 5.98(5)$.

\begin{table}[t]
\begin{center}
\caption{\label{table1} Values for the ladder, crossed, and total contributions to $\bar I$ at various values of $\kappa$.  The numerical integrals were done using Vegas with $50$ iterations of $2G$ function evaluations each.}
\begin{ruledtabular}
\begin{tabular}{cccc}
$\kappa$ & $\bar I_L$ & $\bar I_C$ & $\bar I$ \\
\hline\noalign{\smallskip}
$10$   &   0.744666(5) & -1.009056(5)  & -0.264390(8) \\
$20$   &   0.778244(6) & -1.100290(6)  & -0.322046(9) \\
$30$   &   0.790119(6) & -1,140818(6)  & -0.350800(9) \\
$50$   &   0.799511(7) & -1.180798(6)  & -0.381187(10) \\
$70$   &   0.803522(7) & -1.201271(6)  & -0.397749(10) \\
$100$ &   0.806278(7) & -1.218820(7)  & -0.412542(10) \\
$200$ &   0.809023(7) & -1.243319(7)  & -0.434296(10) \\
$300$ &   0.809689(7) & -1.253166(7)  & -0.443477(10) \\
\end{tabular}
\end{ruledtabular}
\end{center}
\end{table}

As further confirmation of our expressions, we modified our formulas to produce the no-recoil limit (the first term of \eqref{expansion}) directly.  Starting from the heavy particle propagator factor, including uncrossed and crossed photon contributions:
\be \frac{1}{(-q-m_\mu n)^2-m_\mu^2+i \epsilon} + \frac{1}{(-q+m_\mu n)^2-m_\mu^2+i \epsilon} , \label{prop_factor} \ee
we perform the $q^0$ integral using the residue theorem closing the $q^0$ contour in the lower half plane.  The dominant pole has $q^0 \rightarrow \omega-m_\mu-i \epsilon$.  In the no-recoil limit $\omega = \left ( m_\mu^2 + \vec q \,^2 \right )^{1/2} \rightarrow m_\mu$, the propagator factor \eqref{prop_factor} takes the approximate form $i \pi \delta(q^0)/m_\mu$.  We performed first the $q^0$ integral, then the $p$ and $k$ integrals via Feynman parameters, then finally the integral over the magnitude of $\vec q$.  We evaluated the integrals for the ladder and crossed contributions, of six and seven dimensions respectively, using Vegas, and found the numerical results
\bearray \Delta E &=& \left \{ 0.809164(3) - 1.281664(2) \right \} E_F \frac{\alpha^3}{\pi} \cr &=& -0.472500(4) E_F \frac{\alpha^3}{\pi}, \eearray
consistent with the no-recoil result shown as the first term of \eqref{expansion}.


\begin{acknowledgments}
We thank Zvi Bern for an interesting and useful conversation about the evaluation of higher order Feynman diagrams.
\end{acknowledgments}




\end{document}